\documentclass{appolb}

\usepackage{amsmath,amssymb,amsopn}
\usepackage{graphicx}

\newcommand{\be}{\begin{equation}}
\newcommand{\ee}{\end{equation}}
\newcommand{\ba}{\begin{eqnarray}}
\newcommand{\ea}{\end{eqnarray}}
\newcommand{\ban}{\begin{eqnarray*}}
\newcommand{\ean}{\end{eqnarray*}}
\newcommand \nn {\nonumber}

\begin{document}
\title{Parton Energy Loss in Two-Stream Plasma System\thanks{Presented by K. Deja at 
the conference {\it Strangeness in Quark Matter}, Cracow, Poland, September 18-24, 2011.}
}

\author{Margaret E. Carrington
\address{\footnotesize{ Department of Physics, Brandon University,
Brandon, Manitoba, Canada}}
\and
Katarzyna Deja
\address{\footnotesize{National Centre for Nuclear Research, Warsaw, Poland}}
\and
Stanis\l aw Mr\' owczy\' nski
\address{\footnotesize{Institute of Physics, Jan Kochanowski University, Kielce, Poland \\
and National Centre for Nuclear Research, Warsaw, Poland}}
}
\date{January 6, 2012}
\maketitle

\begin{abstract}

The energy loss of a fast parton scattering elastically in a weakly coupled quark-gluon 
plasma is formulated as an initial value problem. The approach is designed to study an 
unstable plasma, but it also reproduces the well known result of energy loss in an equilibrium 
plasma. A two-stream system, which is unstable due to longitudinal chromoelectric
modes, is discussed here some detail. In particular, a strong time and directional 
dependence of the energy loss is demonstrated. 

\end{abstract}

\section{Introduction}

When a highly energetic parton travels through the quark-gluon plasma (QGP), it losses its 
energy due to elastic interactions with plasma constituents. This is the so-called 
{\em collisional energy loss} which for the equilibrium QGP is well understood, see the 
review \cite{Peigne:2008wu} and the handbook \cite{lebellac}. The quark-gluon plasma produced 
in relativistic heavy-ion collisions, however, reaches a state of local equilibrium only after 
a short but finite time interval, and during this period the momentum distribution of plasma 
partons is anisotropic. Collisional energy loss has been computed for such a plasma 
\cite{Romatschke:2004au}  but  the fact that anisotropic QGP evolves fast in time, 
due to chromomagnetic unstable modes (for a review see \cite{Mrowczynski:2005ki}),
has been ignored.

We have developed an approach where the energy loss is found as the solution of an initial 
value problem. The approach is briefly presented in \cite{Carrington:2011uj} where we also 
show that the formalism reproduces the well known result in case of an equilibrium plasma. 
In this paper we discuss in more detail a two-stream system, which is unstable due to longitudinal 
chromoelectric modes, and we demonstrate a strong time and directional dependence of the 
energy loss. 

Throughout the paper we use the natural system of units with $c=\hbar =k_B=1$ and the
signature of our metric tensor is $(+,-,-,-)$.

\section{Energy-loss formula}
\label{general}

We consider a classical parton which moves across a quark-gluon plasma.  Its motion is 
described by the Wong equations \cite{Wong:1970fu}
\ba
\label{EOM-1a}
\frac{d x^\mu(\tau)}{d \tau} &=& u^\mu(\tau ) ,
\\
\label{EOM-1b}
\frac{d p^\mu(\tau)}{d \tau} &=& g Q^a(\tau ) \, F_a^{\mu \nu}\big(x(\tau )\big)
\, u_\nu(\tau ) ,
\\
\label{EOM-1c}
\frac{d Q_a(\tau)}{d \tau} &=& - g f^{abc} u_\mu (\tau ) \,
A^\mu _b \big(x(\tau )\big) \,
Q_c(\tau) ,
\ea
where $\tau$, $x^\mu(\tau )$, $u^\mu(\tau)$ and  $p^\mu(\tau)$ are, respectively, the parton's  
proper time, its trajectory, four-velocity and  four-momentum; $F_a^{\mu \nu}$ and $A_a^\mu$ 
denote the chromodynamic field strength tensor and four-potential along the parton's trajectory
and $Q^a$ is the classical color charge of the parton; $g$ is the coupling constant and
$\alpha_s \equiv g^2/4\pi$ is assumed to be small.  We also assume that the potential 
vanishes along the parton's trajectory  {\it i.e.} our gauge condition is 
$u_\mu (\tau ) \, A^\mu _a \big(x(\tau )\big) = 0 $. Then, according to Eq.~(\ref{EOM-1c}), 
the classical parton's  charge $Q_c(\tau)$ is a constant of motion.

The energy loss is given directly by Eq. (\ref{EOM-1b}) with $\mu = 0$. Using the 
time $t=\gamma\tau$ instead of the proper time $\tau$ and replacing the strength tensor 
$F_a^{\mu \nu}$ by the chromoelectric ${\bf E}_a(t,{\bf r})$ and chromomagnetic 
${\bf B}_a(t,{\bf r})$ fields, Eq.~(\ref{EOM-1b}) gives
\be
\label{e-loss-1}
\frac{dE(t)}{dt} = g Q^a {\bf E}_a(t,{\bf r}(t)) \cdot {\bf v} ,
\ee
where ${\bf v}$ is the parton's velocity. We consider a parton which is very energetic, and therefore
${\bf v}$ is assumed to be constant and ${\bf v}^2 =1$. 

Since we deal with an initial value problem, we apply to the field and current not the usual 
Fourier transformation but  the so-called {\it one-sided Fourier transformation} defined as
\ba
\label{1side}
f(\omega,{\bf k}) &=& \int_0^\infty dt \int d^3r
e^{i(\omega t - {\bf k}\cdot {\bf r})}
f(t,{\bf r}) , \\
\label{1side2}
f(t,{\bf r}) &=& \int_{-\infty+i\sigma}^{{\infty+i\sigma}} \frac{d\omega}{2\pi} \int \frac{d^3k}{(2\pi)^3}
e^{-i(\omega t - {\bf k}\cdot {\bf r})}
f(\omega,{\bf k}) ,
\ea
where the real parameter $\sigma > 0$ is chosen is such a way that the integral over $\omega$ 
is taken along a straight line in the complex $\omega-$plane, parallel to the real axis, above all 
singularities of $f(\omega,{\bf k})$. Introducing the current generated by 
the parton ${\bf j}_a(t,{\bf r}) = g Q^a {\bf v} \delta^{(3)}({\bf r} - {\bf v}t)$ and using Eq. (\ref{1side2}), 
Eq.~(\ref{e-loss-1}) can be rewritten:
\be
\label{e-loss-3}
\frac{dE(t)}{dt} = g Q^a
\int_{-\infty +i\sigma}^{\infty +i\sigma}
{d\omega \over 2\pi}
\int {d^3k \over (2\pi)^3}
e^{-i(\omega - \bar\omega)t} \; {\bf E}_a(\omega,{\bf k}) \cdot {\bf v} ,
\ee
where $\bar\omega \equiv {\bf k} \cdot {\bf v}$. 

The next step is to compute the chromoelectric field ${\bf E}_a$. Applying the one-sided 
Fourier transformation to the linearized Yang-Mills equations, we get the chromoelectric
field given as
\be
\label{E-field-k}
E^i_a(\omega, {\bf k}) = -i
(\Sigma^{-1})^{ij}(\omega,{\bf k})
\Big[ \omega j_a^j(\omega,{\bf k})
+ \epsilon^{jkl} k^k B_{0a}^l ({\bf k})
- \omega D_{0a}^j ({\bf k}) \Big] ,
\ee
where  $B_0$ and $D_0$ are the initial values of the chromomagnetic field and  
the chromoelectric induction 
$D^i_a(\omega, {\bf k}) = \varepsilon^{ij}(\omega, {\bf k}) E^j_a(\omega, {\bf k})$. 
The chromodielectric tensor $\varepsilon^{ij}(\omega, {\bf k})$ equals
\begin{equation*}
\varepsilon^{ij} (\omega,{\bf k})
=  \delta^{ij} + 
{g^2 \over 2 \omega} \int {d^3 p \over (2\pi )^3}
{ v^i \over \omega - {\bf k} \cdot {\bf v} + i0^+} 
{\partial f({\bf p}) \over \partial p^k} 
\Big[ \Big( 1 - {{\bf k} \cdot {\bf v} \over \omega} \Big) \delta^{kj}
+ {k^k v^j \over \omega} \Big] ,
\end{equation*}
where $ f({\bf p})$ is the momentum distribution of plasma constituents. The color indices $a,b$ 
are dropped because $\varepsilon (\omega, {\bf k})$ is a unit matrix in color space. The matrix 
$\Sigma^{ij}(\omega,{\bf k})$ from Eq.~(\ref{E-field-k}) is defined
\be
\label{matrix-sigma}
\Sigma^{ij}(\omega,{\bf k}) \equiv
- {\bf k}^2 \delta^{ij} + k^ik^j
+ \omega^2 \varepsilon^{ij}(\omega,{\bf k}) .
\ee
Substituting the expression (\ref{E-field-k}) into Eq.~(\ref{e-loss-3}), we obtain the formula
\ba
\label{e-loss-6}
\frac{dE(t)}{dt} &=& g Q^a v^i \int_{-\infty +i\sigma}^{\infty +i\sigma}
{d\omega \over 2\pi i}
\int {d^3k \over (2\pi)^3}
e^{-i(\omega -\bar{\omega})t}
(\Sigma^{-1})^{ij}(\omega,{\bf k})
\\ [2mm]\nn
&\times&
\Big[ 
\frac{i \omega g Q^a v^j}{\omega - \bar{\omega}}
+ \epsilon^{jkl} k^k B_{0a}^l ({\bf k})
- \omega D_{0a}^j ({\bf k}) \Big] .
\ea

When the plasma is stable, the poles of $\Sigma^{-1}(\omega,{\bf k})$ 
are located in the lower half-plane of complex $\omega$ and the corresponding 
contributions to the energy loss from these poles decay exponentially in time.
The only stationary contribution to the energy loss is  given by the pole 
$\omega = \bar{\omega} \equiv {\bf k}\cdot {\bf v}$ from the first term in square brackets 
in Eq.~(\ref{e-loss-6}). Therefore, the terms  which depend on the initial values of the fields, 
can be neglected, and Eq.~(\ref{e-loss-6}) reproduces the known result for the energy-loss 
of a highly energetic parton in an equilibrium quark-gluon plasma \cite{Carrington:2011uj}.

When the plasma is unstable, the matrix $\Sigma^{-1}(\omega,{\bf k})$ contains poles 
in the upper half-plane of complex $\omega$, and the contributions to the energy loss from 
these poles grow exponentially in time. In this case, the terms in Eq.~(\ref{e-loss-6}) which 
depend on the initial values of the fields ${\bf D}$ and ${\bf B}$ cannot be neglected. Using 
the linearized Yang-Mills equations, the initial values $B_0$ and $D_0$ are 
expressed through the current and we obtain
\ba
\label{e-loss-unstable}
&& \frac{d\overline{E}(t)}{dt} = g^2 C_R 
v^i v^l \int_{-\infty +i\sigma}^{\infty +i\sigma}
{d\omega \over 2\pi}
\int {d^3k \over (2\pi)^3}
e^{-i(\omega - \bar{\omega}) t}
(\Sigma^{-1})^{ij}(\omega,{\bf k})
\\ \nn
&&\times
\Big[ 
\frac{\omega \delta^{jl}}{\omega - \bar{\omega}}
-(k^j k^k - {\bf k}^2 \delta^{jk})
(\Sigma^{-1})^{kl}(\bar{\omega},{\bf k}) 
+ \omega \, \bar{\omega} \, \varepsilon^{jk}(\bar{\omega},{\bf k})
(\Sigma^{-1})^{kl}(\bar{\omega},{\bf k}) ,  
 \Big] 
\ea
where $C_R$ comes from the averaging over color states of the test parton, $C_R = 4/3$ for 
a quark and $C_R = 3$ for a gluon. 

\section{Two-stream system}
\label{sec-2-streams}

In order to calculate the energy loss, one must invert the matrix $\Sigma^{ij}(\omega,{\bf k})$ 
defined by Eq.~(\ref{matrix-sigma}) and substitute the resulting expression into 
Eq. (\ref{e-loss-unstable}). For a general anisotropic system this is a tedious calculation.
In the case of the two-stream system, which has unstable longitudinal electric modes, the 
chromodynamic field is dominated after a sufficiently long time by the longitudinal chromoelectric 
component. Consequently, we assume that ${\bf B}(\omega, {\bf k}) = 0$ and  
${\bf E}(\omega, {\bf k}) = {\bf k} \big({\bf k}\cdot {\bf E}(\omega, {\bf k})\big) / {\bf k}^2$, in which 
case the tensor $\Sigma^{ij}(\omega,{\bf k})$ is trivially inverted as
\be
\label{inv-sigma}
(\Sigma^{-1})^{ij}(\omega,{\bf k}) = 
\frac{1}{\omega^2 \varepsilon_L(\omega,{\bf k})}
\frac{k^ik^j}{{\bf k}^2}\,,~~
\varepsilon_L(\omega,{\bf k}) \equiv
\varepsilon^{ij}(\omega,{\bf k}) \frac{k^ik^j}{{\bf k}^2},
\ee
and Eq. (\ref{e-loss-unstable}) simplifies to 
\ba
\label{e-loss-2-stream-1}
\frac{d\overline{E(t)}}{dt} &=& 
 g^2 C_R 
\int_{-\infty +i\sigma}^{\infty +i\sigma}
{d\omega \over 2\pi}
\int {d^3k \over (2\pi)^3}
\frac{e^{-i(\omega - \bar{\omega}) t}}
{\omega^2 \varepsilon_L(\omega,{\bf k})}
\frac{\bar{\omega}^2}{{\bf k}^2}
\Big[ 
\frac{\omega}{\omega - \bar{\omega}}
+   \frac{\bar{\omega}}{\omega}
\Big] .
\ea
Eq.~(\ref{e-loss-2-stream-1}) gives a non-zero energy loss in the vacuum limit when 
$\varepsilon_L \rightarrow 1$.  Therefore, we subtract from the formula (\ref{e-loss-2-stream-1}) 
the vacuum contribution, or equivalently we replace $1/\varepsilon_L$ by  $1/\varepsilon_L - 1$.

The next step is to calculate $\varepsilon_L(\omega,{\bf k})$. With the distribution 
function of the two-stream system in the form 
$
f({\bf p}) = (2\pi )^3 n 
\Big[\delta^{(3)}({\bf p} - {\bf q}) + \delta^{(3)}({\bf p} + {\bf q}) \Big] ,
$ where $n$ is the effective parton density in a single stream, one finds \cite{Mrowczynski:2008ae} 
\be
\label{eL-2-streams}
\varepsilon_L(\omega,{\bf k}) 
= \frac{\big(\omega - \omega_+({\bf k})\big)
\big(\omega + \omega_+({\bf k})\big)
\big(\omega - \omega_-({\bf k})\big)
\big(\omega + \omega_-({\bf k})\big)}
{\big(\omega^2 - ({\bf k} \cdot {\bf u})^2\big)^2} ,
\ee
where ${\bf u} \equiv {\bf q}/E_{\bf q}$ is the stream velocity, and $\mu^2 \equiv g^2n/2 E_{\bf q}$ 
is a parameter analogous to the Debye mass squared. There are four roots to the dispersion relation 
$\varepsilon_L(\omega,{\bf k}) = 0$ which read
\ba
\label{roots}
\omega_{\pm}^2({\bf k}) &=& \frac{1}{{\bf k}^2}
\Big[{\bf k}^2 ({\bf k} \cdot {\bf u})^2
+ \mu^2 \big({\bf k}^2 - ({\bf k} \cdot {\bf u})^2\big)
\\[2mm] \nonumber
&\pm& \mu \sqrt{\big({\bf k}^2 - ({\bf k} \cdot {\bf u})^2\big)
\big(4{\bf k}^2 ({\bf k} \cdot {\bf u})^2 +
\mu^2 \big({\bf k}^2 - ({\bf k} \cdot {\bf u})^2\big)\big)} 
\; \Big] .
\ea
It is easy to see that $0 < \omega_+({\bf k}) \in \mathbb{R}$ for any ${\bf k}$.  
For ${\bf k}^2 ({\bf k} \cdot {\bf u})^2 \ge 2 \mu^2 \big({\bf k}^2 - ({\bf k} \cdot {\bf u})^2\big)$
the minus mode is also stable, $0 < \omega_-({\bf k}) \in \mathbb{R}$. However, for  
${\bf k} \cdot {\bf u} \not= 0$ and 
${\bf k}^2 ({\bf k} \cdot {\bf u})^2 < 2 \mu^2 \big({\bf k}^2 - ({\bf k} \cdot {\bf u})^2\big)$ one finds 
$\omega^2_-({\bf k})<0$ and $\omega_-({\bf k})$  imaginary. This is the well-known two-stream 
electric instability. 

Strictly speaking, the stream velocity ${\bf u}$ given by the distribution function equals the 
speed of light. However, the distribution function should be treated as an idealization of 
a two-bump distribution with bumps of finite width. This means that the momenta 
of all partons are not  exactly parallel or antiparallel, and the velocity ${\bf u}$ which enters 
Eqs.~(\ref{eL-2-streams}, \ref{roots}) obeys ${\bf u}^2 \le 1$. 

Equations (\ref{e-loss-2-stream-1}) and (\ref{eL-2-streams}) determine the energy loss of 
a parton in the two-stream system.  The integral over $\omega$ can be computed analytically 
as it is determined by the six poles of the integrand located at  
$\omega = \pm \omega_+({\bf k}), \; \pm \omega_-({\bf k})$, $\bar\omega$ and $0$. The 
remaining integral over ${\bf k}$ must be done numerically. We use cylindrical coordinates
(${\bf k} = (k_T, \phi,k_L)$) with the axis $z$ along the streams.  The parameters are chosen to be:
 $g=|{\bf v}| = 1$, $|{\bf u}| = 0.9$, $C_R = 3$. As discussed in \cite{Carrington:2011uj},  
the integral over ${\bf k}$ is divergent and it has been taken over a domain such that  
$-k_{\rm max} \le k_L \le k_{\rm max}$ and  $0 \le k_T \le k_{\rm max}$ with $k_{\rm max} = 20\mu$. 

\begin{figure}[t]
\centering
\includegraphics[width=0.9\textwidth]{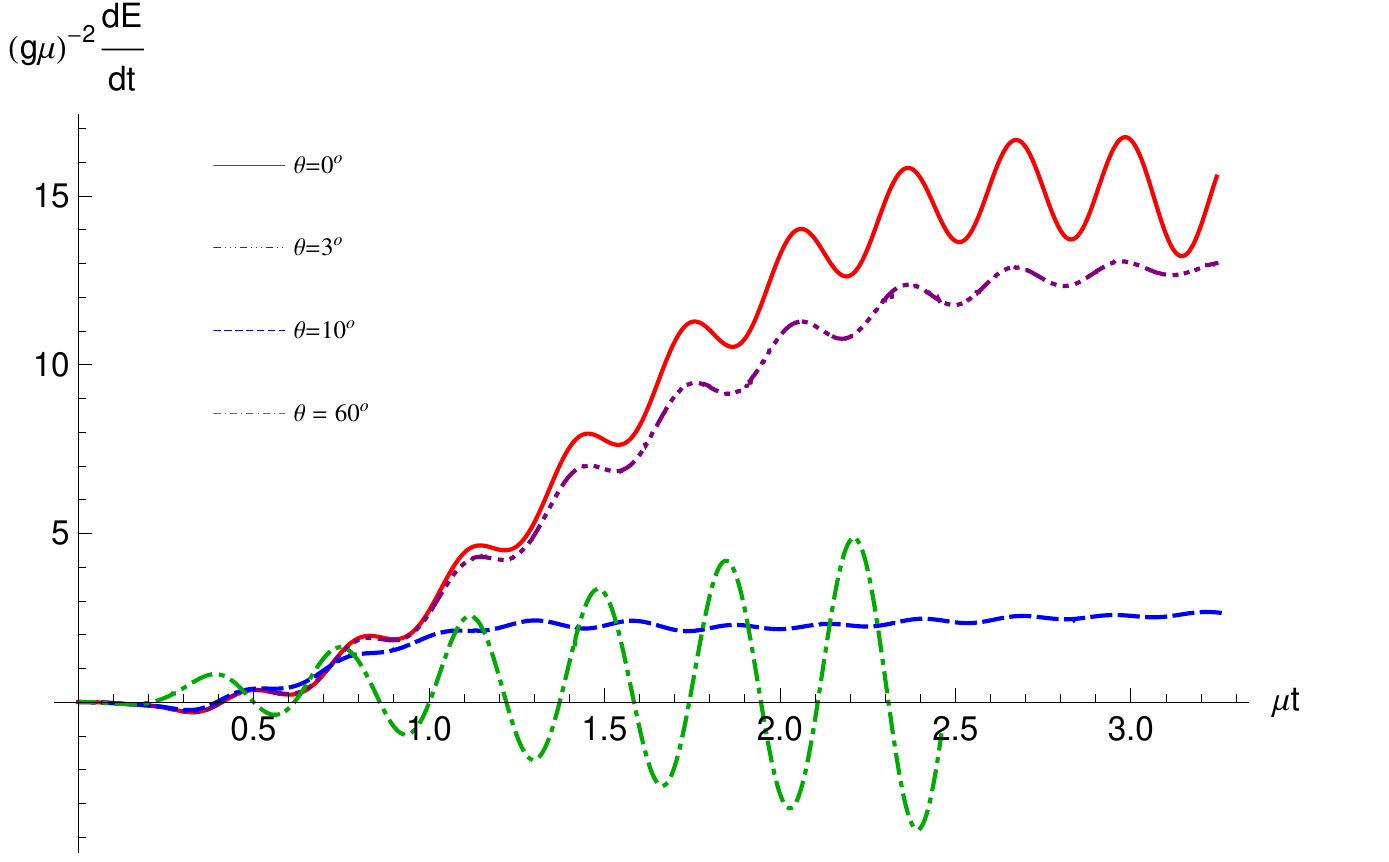}
\caption{The parton energy loss per unit length as a function of time for several angles 
$\theta$ between the parton's velocity ${\bf v}$ and stream velocity ${\bf u}$.}
\label{fig-e-loss-directional}
\end{figure}

In Fig.~\ref{fig-e-loss-directional} we show the parton energy loss per unit length as a function of time 
for different orientations of the parton's velocity  ${\bf v}$ with respect to the stream velocity ${\bf u}$.  
The energy loss oscillates and manifests a strong directional dependence.

\section{Conclusions}

We have developed a formalism where the energy loss of a fast parton in a plasma medium
is found as the solution of an initial value problem. The formalism determines the energy
loss in an unstable plasma which contains modes that exponentially grow in time. The two-stream 
system has been studied in some detail. The energy loss per unit length is not constant, as in 
an equilibrium plasma, but it exhibits strong time and directional dependences.

\section*{Acknowledgments}

This work was partially supported by Polish Ministry of Science and Higher Education under 
grants N~N202~204638 and 667/N-CERN/2010/0.



\begin{thebibliography}{99}

\bibitem{Peigne:2008wu}
S.~Peigne and A.~V.~Smilga,
Phys.\ Usp.\  {\bf 52}, 659 (2009).

\bibitem{lebellac}
M.~ Le~Bellac, 
{\it Thermal Field Theory}  
(Cambridge University Press, Cambridge, 2000).

\bibitem{Romatschke:2004au}
P.~Romatschke and M.~Strickland,
Phys.\ Rev.\  D {\bf 71}, 125008 (2005).

\bibitem{Mrowczynski:2005ki}
St.~Mr\'owczy\'nski,
Acta Phys.\ Polon.\  B {\bf 37}, 427 (2006).

\bibitem{Carrington:2011uj} 
M.~E.~Carrington, K.~Deja and St.~Mr\'owczy\'nski,
arXiv:1110.4846 [hep-ph].  

\bibitem{Wong:1970fu}
S.~K.~Wong,
Nuovo Cim.\  A {\bf 65}, 689 (1970).

\bibitem{Mrowczynski:2008ae}
St.~Mr\'owczy\'nski,
Phys.\ Rev.\  D {\bf 77}, 105022 (2008).


\end{thebibliography}
\end{document}